\documentclass[%
 reprint,
prb,
]{revtex4-2}

\usepackage{graphicx}
\usepackage{dcolumn}
\usepackage{bm}

\usepackage[utf8]{inputenc}
\usepackage[T1]{fontenc}
\usepackage{mathptmx}

\begin{document}

\preprint{APS/123-QED}

\title[An \textsf{achemso} demo]
  {Ultrahigh-Frequency Wireless MEMS QCM Biosensor for Direct Label-Free Detection of Biomarkers in a Large Amount of Contaminants}

\author{Kentaro Noi$^{1}$}
\author{Arihiro Iwata$^{2}$}
\author{Fumihito Kato$^{3}$}
\author{Hirotsugu Ogi$^{1*}$}
\affiliation{$^{1}$Graduate School of Engineering, Osaka University, Suita, Osaka 565-0871, Japan}
\affiliation{$^{2}$Graduate School of Engineering Science, Osaka University, Toyonaka, Osaka 560-8531, Japan}
\affiliation{$^{3}$Department of Mechanical Engineering, Nippon Institute of Technology, Saitama 345-8501, Japan}
\email[]{ogi@prec.eng.osaka-u.ac.jp}


\begin{abstract}
Label-free biosensors, including conventional quartz-crystal-microbalance (QCM) biosensor, are seriously affected by nonspecific adsorption of contaminants involved in analyte solution, and it is exceptionally difficult to extract the sensor responses caused only by the targets.  In this study, we reveal that this difficulty can be overcome with an ultrahigh-frequency wireless QCM biosensor.  The sensitivity of a QCM biosensor dramatically improves by thinning the quartz resonator, which also makes the resonance frequency higher, causing high-speed surface movement.  Contaminants weakly (nonspecifically) interact with the quartz surface, and they fail to follow the fast surface movement and cannot be detected as the loaded mass.  The targets are, however, tightly captured by the receptor proteins immobilized on the surface, and they can move with the surface, contributing to the loaded mass and decreasing the resonant frequency.  We develop a MEMS QCM biosensor, in which an AT-cut quartz resonator of 26 $\mu$m thick is packaged without fixing, and demonstrate this phenomenon by comparing the frequency changes of fundamental ($\sim$64 MHz) and ninth ($\sim$576 MHz) modes.  At ultrahigh-frequency operation with the ninth mode, the sensor response is independent of the amount of impurity proteins, and the binding affinity is unchanged.  We then applied this method for the label-free and sandwich-free direct detection of C-reactive protein (CRP) in serum, and confirmed its applicability.\\
\\
Analytical Chemistry 91(15), 9398-9402 (2019)\\
https://doi.org/10.1021/acs.analchem.9b01414
\end{abstract}

\maketitle

\section{INTRODUCTION}
Label-free biosensors allow us to detect and quantify target biomolecules in analyte solution in a short time, because they do not require any incubation and washing processes.  Also, they are capable of monitoring binding and dissociation reactions between biomolecules, which provides us with binding affinity between them.  Many label-free biosensors have been proposed, including the quartz-crystal-microbalance (QCM) biosensor \cite{Muramatsu, Liu, Ogi2006, Patel2009}, the surface-plasmon-resonance (SPR) biosensor \cite{Liedberg, Nedelkov, Hoa}, the optical resonator biosensor \cite{Ramachandran, Luchansky, Ren}, the thermal-phonon biosensor \cite{Peeters, Ogi2019} and so on.   

However, a key issue remains unsolved; the direct label-free assay in solution involving a large amount of contaminants.  Concentration of a target biomolecule is usually much lower than those of other impurity proteins in practical analyte solutions such as serum, plasma, and urea, and those contaminants nonspecifically adsorb on the sensor surface, producing signifiant (apparent) sensor response and making the response from the target undetectable.  This effect becomes signifiant even when the affinity between the contaminants and the sensor surface is lower because of their much larger concentrations.  For example, C-reactive-protein (CRP), known as a biomarker of inflammation, is used for evaluating cardiovascular disease, and a CRP level higher than 3 $\mu$g/ml indicates high risk \cite{Ridker}, whereas concentration of albumin in blood is about 50 mg/ml, higher than the CRP concentration by a factor of 17,000.  Therefore, detection of biomarkers in such unclean solutions has been performed by a sandwich method, which consists of several steps; injection of analyte and incubation, washing the sensor chip to remove the nonspecifically bonded contaminants, and injection of the detection antibodies to quantify the target materials. In QCM experiments, for example, the mass-amplified sandwich assays were performed for high sensitive detection of biomarkers \cite{Kim, Ogi2011, Ding}.  However, the sandwich assay requires several extra procedures and deteriorates the important advantage of the label-free biosensor, that is, the short-time and real-time assay.  

We here propose ultrahigh-frequency (UHF) QCM measurement to overcome this difficulty.  The QCM biosensor detects target materials captured by surface-immobilized receptors through addition of mass; it is therefore the mass-sensitive biosensor.  The attached targets are sensed as added mass when they move together with the quartz crystal, so that the resonance frequency decreases because of enhanced inertia resistance.  However, when the surface movement is too fast for adsorbed materials to follow the movement, they cannot be detected as added mass.  The antigen-antibody binding is highly strong and the captured target by its antibody will follow the surface movement.  On the other hand, weakly bonded impurity materials will fail to follow the fast surface movement at UHF region and fail to contribute the resonance-frequency change.  Therefore, it is possible to directly detect the target materials with UHF QCM measurements even in a solution with a large amount of contaminants (Fig. 1).  It is, however, never straightforward to realize a UHF QCM biosensor with conventional methodology with electrodes and wires on the quartz resonator; heavy electrodes of noble metals and attached wires significantly restrict vibration at high frequencies in solution \cite{Ogi2009}, and we have proposed wireless-electrodeless QCM biosensors, which allowed QCM biosensor with frequencies much higher than conventional QCM systems \cite{Ogi2007, Ogi2011}.  The wireless QCM biosensor was then fabricated by using micro-electro mechanical system (MEMS) technique, where a thin AT-cut quartz resonator was packaged between micropillars in the silicon microchannel without fixing the quartz resonator \cite{Kato2011, Kato2012}, which allowed QCM experiments at $\sim$400 MHz \cite{Shagawa}. 

In this study, we fabricate an improved wireless MEMS QCM biosensor with fundamental resonance frequency of 64 MHz, which operates at frequencies up to ninth mode ($\sim$576 MHz) for the label-free direct detection of biomarkers in solutions with many contaminants.  The number of micorpillars is increased for achieving higher stability at UHF range and for stirring the solution.  We also develop a high-power analog-based heterodyne spectrometer for the non-contacting measurement of the resonance frequency.  The model biomarker proteins are rabbit immunoglobulin G (rIgG) and CRP, and the model contaminant is bovine serum albumin (BSA).  We first investigate influence of BSA on the binding affinity between rIgG and protein A and that between CRP and anti-CRP antibody at fundamental and ninth modes; we confirm that the binding affinity is identical to that without the contaminants at the UHF mode.  We then demonstrate the ability of the UHF QCM experiment by directly detecting CRP in serum.

\section{MATERIALS AND METHODS}
\subsection{Wireless MEMS QCM system}
Figure S1 in Supplementary Information illustrates the structure of the wireless MEMS QCM developed in this study.  It consists of three layers; top and bottom glass layers and the middle Si layer.  The microchannels were fabricated at the top and middle layers, where the AT-cut quartz resonator was packaged.  The micropillars were prepared at the top glass layer and the middle Si layer, between which the quartz resonator was embedded without fixing.  The anodic bonding technique was used to tightly bond glass and Si.  The details of the MEMS process appear in our previous studies \cite{Kato2011, Kato2012}.  The number of micropillars is increased to 59 from 4 of the previous design \cite{Kato2012}.   5 micropillars are located before and after the quartz resonator for stirring the solution flow and 49 micropillars support the resonator. 
\begin{figure}[ht]
\begin{center}
\includegraphics[width=84mm]{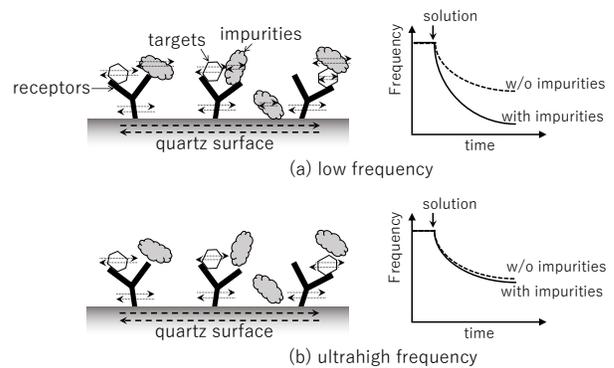}
\end{center}
\caption{Schematics of (a) low-frequency and (b) ultrahigh-frequency (UHF) QCM experiments.  In the former, impurities adsorbed nonspecifically on the quartz surface vibrate following the surface movement as well as targets, causing excess frequency change.  In the latter, the weakly attached impurities cannot vibrate with the surface, and they fail to affect the frequency change.}
\label{}
\end{figure}

We attached copper tapes on the top and bottom faces of the MEMS QCM chip as excitation and detection antennas for generating and detecting the shear vibration of the quartz resonator, respectively.  The MEMS QCM chip was then set it in a sensor cell to make the solution flow as shown in Fig. S2 in Supplementary Information.  A piezoelectric micropump (Takasago Electric, Inc. model No. AM-152-2) was used for flowing solutions with a flow rate of 100 $\mu$l/min.
  
Excitation and detection of the shear vibrations of the resonator were performed by the UHF heterodyne spectrometer developed in this study.  Its details appear in Fig. S3 in Supplementary Information.  Tone bursts of $\sim$5 $\mu$s were applied to the generation antenna, which launched the electromagnetic field and caused the shear vibration of the resonator through the inverse piezoelectric effect.  After the excitation, the resonator continued to vibrate with its resonance frequency, generating the electromagnetic field through the piezoelectric effect.  This reverberating signal was detected by the detection antenna and entered the analog heterodyne spectrometer, where the amplitude and phase at the driving frequency component were extracted.  The analog electronics is based on that developed previously \cite{OgiRF2009}, but frequency of each component is increased.  The resonance spectrum was obtained by sweeping the driving frequency and obtaining the amplitudes, and the resonance frequency was determined from the peak frequency.  In the QCM measurement, we fixed the diver frequency at the baseline resonance frequency and monitored the change in the phase of the reverberating signal, which was converted to the frequency change \cite{Ogi2009}.  

\subsection{Preparation of resonator chip}
Before packaging the quartz resonator inside the microchannel by the anodic bonding method, we deposited 5 nm Cr and then 15 nm Au thin films on both surfaces.  After the boding procedure, we injected a piranha solution (98\% H$_2$SO$_4$ : 33\% H$_2$O$_2$=7:3) for cleaning the resonator and rinsed it with ultrapure water.  For making the self-assembled monolayers on both surfaces, we injected a 10-mM 10-carboxy-decanthiol solution with absolute-ethanol buffer and incubated for 12 h at 4 $^\circ$C.  After rinsing the resonator with absolute ethanol and then ultrapure water, we injected a 100-mM EDC (1-ethyl-3- (3-dimethylaminopropyl)carbodiimide, hydrochloride) solution with 100mM NHS (N-hydroxysulfosuccinimide sodium salt) in ultrapure water and incubated for 1 h at 20 $^\circ$C for activating the sensor surfaces.  After rising the resonator with ultrapure water, the receptor protein solutions (shown below) were injected and incubated for immobilizing receptor proteins on both surfaces.  Finally, we injected a 10 mg/ml BSA solution and incubated for 1 h for blocking remaining activated ester sites.  

We used two target model proteins in this study.  One is rIgG and the other is CRP, and they are diluted with phosphate buffered saline (PBS) solution with or without BSA.  We also prepared CRP solutions using fetal bovine serum (FBS) for simulating practical sample.  The receptor protein solution is either a 100-$\mu$g/ml protein A solution for detecting rIgG or a 100-$\mu$g/ml anti-CRP antibody solution for detecting CRP.  The time and temperature in immobilizing receptor proteins are 1 h at 20 $^\circ$C for protein A and 3 h at 25 $^\circ$C for anti-CRP antibody.

CRP (\#8C72) and anti-CRP antibody (\#4C28-CRP135) were obtained from HyTest Ltd.  rIgG (\#148-09551) and NHS (\#087-09371) were obtained from Wako.  Protein A (\#101100) was from Invtrogen.  BSA (\#A3059) was from Sigma Aldrich.  10-Carboxy-decanthiol (\#C385) and EDC (\#W001) were from DOJINDO.  FBS (\#10270-106) was from Gbico.

\section{RESULTS}
Figure 2(a) shows resonance-frequency changes of the fundamental mode (64 MHz) when 6.7 nM rIgG solutions containing various amounts of BSA were injected.  It is clearly seen that the frequency change is significantly affected by the contaminants, indicating the difficulty in evaluating target concentration and target-receptor kinetics accurately.  On the other hand, the frequency changes are unaffected by the contaminants in the UHF experiments as shown in Fig. 2(b).
\begin{figure}[ht]
\begin{center}
\includegraphics[width=84mm]{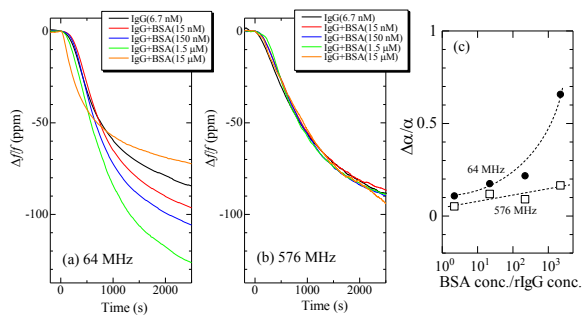}
\end{center}
\caption{Binding curves in injection of 6.7-nM rIgG solutions containing various concentration BSA in PBS measured at (a) 64 MHz and (b) 576 MHz.  (c) Change of the exponential coefficient of the frequency change caused by the BSA contaminant.  (The horizontal axis indicates the concentration ratio between target and contaminant.)}
\label{Model}
\end{figure}

\begin{figure*}[t]
\begin{center}
\includegraphics[width=150mm]{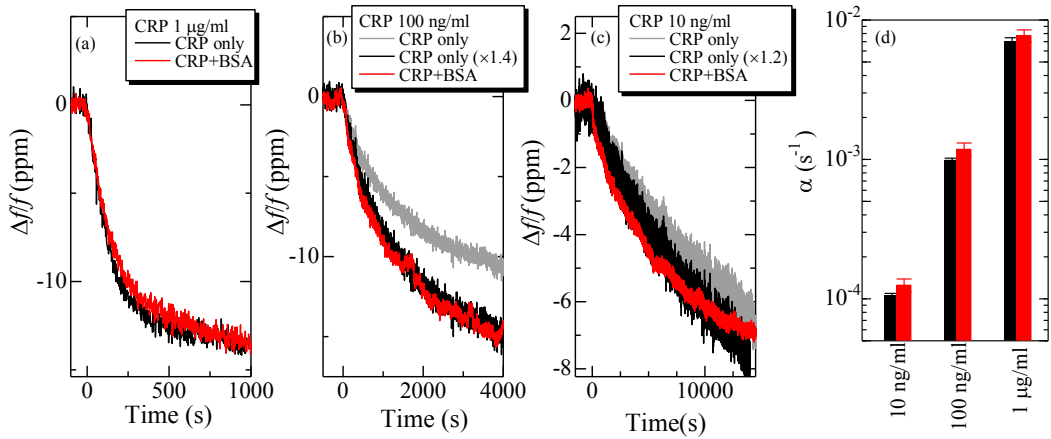}
\end{center}
\caption{Binding curves in detecting (a) 1000, (b) 100, and (c) 10 ng/ml CRP in PBS solutions measured at 576 MHz.  Red and black lines are measurements for solutions with and without 15 $\mu$M BSA, respectively.  The gray lines in (b) and (c) are as-measured data; the black lines are their amplified data for emphasizing the identical exponential coefficients in the frequency change with the BSA contaminant.  (d) Comparison of the exponential coefficients of the frequency change between with (red) and without (black) 15 $\mu$M BSA in the three concentration CRP solutions.}
\end{figure*}

Figures 3(a)-(c) compare the frequency changes in detecting CRP between with and without BSA for CRP concentrations of 10, 100, and 1000 ng/ml in the UFH QCM experiment.  We again observe identical frequency changes between with and without the contaminants.  

Figure 4(a) shows the frequency change in detecting CPR in the FBS solution.  Injection of the CRP-in-FBS solution caused a baseline jump, after which the resonance frequency decreased.  Figure 4(b) shows the relationship between the total frequency decrease and CRP concentration in FBS.
\begin{figure}[t]
\begin{center}
\includegraphics[width=84mm]{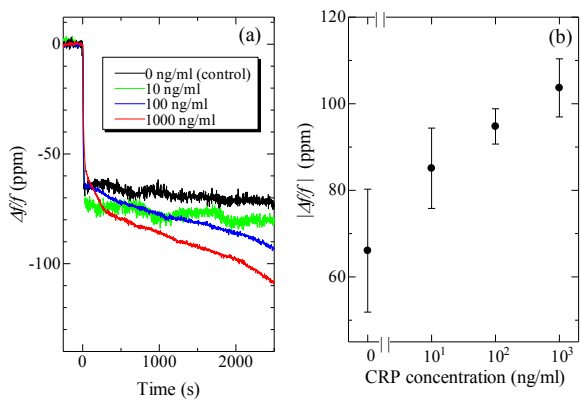}
\end{center}
\caption{(a) Binding curves measured at 576 MHz in detecting CRP in fetal bovine serum (FBS).  (b) Relationship between the frequency change at 2500 s and the CRP concentration in FBS.}
\end{figure}

\section{DISCUSSIONS}
First, we discuss the mass sensitivity of our QCM biosensor.  The mass sensitivity of a QCM is evaluated by \cite{OgiPJAS}
\begin{equation}
\left| \frac{\Delta f_1}{\rho_s}\right|=\frac{A_e v_q}{2A_q\rho_q}\cdot \frac{1}{d_q},
\end{equation}
where $\Delta f_1$ denotes the change in the fundamental resonance frequency, $\rho_s$ the area mass density (adsorbed mass per unit area), $A_q$ the one-sided surface area of the resonator, $A_e$ the effective reaction area on which targets accumulate, $v_q$ the shear wave velocity in AT-cut quartz, and $d_q$ the resonator thickness.  $A_e/A_q$ value becomes smaller than unity in conventional QCMs because a part of single surface is used for sensing, whereas our MEMS QCM can use both surfaces, so that $A_e/A_q$ value can be the maximum of two.  The resonator thickness of our MEMS QCM (26 $\mu$m) is significantly smaller than those of conventional 10 or 5 MHz QCMs ($>\sim$160 $\mu$m or $>\sim$330 $\mu$m, respectively).  The mass sensitivity $\left|\Delta f_1 / \rho_s\right|$ of our MEMS QCM (=18.4 Hz$\cdot$ cm$^2$/ng) is, therefore, larger than those of the conventional QCMs ($\sim$0.2 or less) by a factor of 90 or more.

It is important to mention that even such a high-frequency (64 MHz) QCM fails to detect the target proteins in the BSA contaminants.  Figure 2(a) shows that the amount of the frequency change increases as the BSA molecule increases up to 1.5 $\mu$M, which we attribute to the nonspecific adsorption of BSA molecules on the sensor surfaces.  The amount of the frequency change decreases at the maximum contaminants (15 $\mu$M), but frequency change occurs rapidly. 
This can be explained by the kinetics in the flow-injection system: According to the kinetic theory under the pseudo-first-order reaction\cite{Eddowes} and the Sauerbrey equation \cite{Sauerbrey}, the frequency change obeys an exponential function ($\Delta f\propto (1-\textrm{e}^{-\alpha t})$)\cite{Liu, OgiNonspe} and its exponential coefficient $\alpha$ is expressed by
\begin{equation}
\alpha=k_aC_{t}+k_d,
\end{equation}
where $k_a$ and $k_d$ are the reaction velocity constants for association and dissociation, respectively, and $C_{t}$ denotes the concentration of target.  Excess contaminants will apparently increase $C_{t}$ and then $\alpha$, causing a steep frequency change.  As shown in Fig. 2(c), the $\alpha$ value apparently increases with the increase of the amount of BSA molecules.  Therefore, quantification of the target protein becomes significantly difficult even with the high-frequency (64 MHz) QCM.  However, using the UHF QCM, the frequency response is almost insensitive to the amount of BSA even with BSA concentration of 15 $\mu$M (Fig. 2(b)), which is larger than the target concentration by a factor of 2,238.  The exponential coefficient, $\alpha$, is also nearly unaffected by the contaminants as shown in Fig. 2(c). 
 
We evaluated the dissociation constant $K_D=k_d/k_a$ with and without the BSA contaminants by injecting different concentration rIgG solutions and measuring the corresponding $\alpha$
values \cite{Ogi2008}.  The $K_D$ value between rIgG and protein A is 2.0 nM without BSA, but this value increased to 12 nM in solution containing 15 $\mu$M BSA when we use the fundamental mode (64 MHz): Thus, the binding affinity is apparently lowered significantly by the presence of BSA, because BSA molecules nonspecifically bind to the surface-immobilized protein A and prevent rIgG molecules from interacting with protein A.  However, when we measure the $K_D$ value at 576 MHz with 15 $\mu$M BSA, it become 1.7 nM, being identical to that without BSA. 

Similar experiments were performed in detecting CRP.  As shown in Fig. 3(a)-(c), the binding curves are unaffected by the presence of BSA contaminants by using 576-MHz resonance mode, and their exponential coefficients are also nearly the same between with and without 15 $\mu$M BSA (Fig. 3(d)).  The result in detecting 10 ng/ml CRP (Fig. 3(c)) is notably remarkable because the molar concentration of the BSA contaminant exceeds that of the target by a factor of 150,000.  Furthermore, the $K_D$ value between CRP and anti-CRP antibody was 0.23 nM without BSA and was 0.29 nM with 15 $\mu$M BSA; they well agree with each other.  These experiments and analysis highly confirm the validity of our proposal that UFH QCM experiment allows the label-free direct detection of target proteins in many contaminants.

Results in Fig. 4 indicate high applicability of this methodology to practical applications.  We always observed a frequency jump of about 65 ppm at the arrival of the CRP-in-FBS solution, which will be caused by higher viscosity of FBS.  However, we succeeded in observing clear difference in the frequency change after the baseline jump.  As shown in Fig. 4(b), it is possible to detect 10 ng/ml CRP within $\sim$40 min even in FBS without any sandwich approach nor any reference channels.       

\section{CONCLUSIONS}
We developed a MEMS QCM biosensor system with a 26-$\mu$m thick AT-cut quartz resonator, and investigated influence of contaminants in solution in detecting target proteins.  We used rIgG and CRP for the target proteins, and protein A and anti-CRP antibody for their receptors, respectively.  By mixing BSA molecules up to 15 $\mu$M, we measured the binding curves at the fundamental mode (64 MHz) and ninth mode (576 MHz).  The binding curves were significantly affected by the presence of BSA in the experiments at 64 MHz, highly lowering the binding affinity.  However, we observed identical binding curves, being independent of the BSA concentration in the experiments at 576 MHz, yielding to the identical dissociation constants.  Furthermore, this UHF MEMS QCM was applied to perform the direct detection of CRP in serum and succeeded in detecting 10 ng/ml CRP in serum without using any sandwich assay.



\section*{acknowledgement}
This study was supported by the Development of Advanced Measurement and Analysis Systems from Japan Science and Technology Agency, JST.

\end{document}